\documentclass[twocolumn,showpacs,preprintnumbers,amsmath,amssymb]{revtex4}

\usepackage{graphicx}
\usepackage{dcolumn}
\usepackage{bm}

\begin{document}

\title{Phase transition in site-diluted Josephson-junction arrays: A numerical study}
\author{Jian-Ping Lv$^{1}$, Huan Liu$^{1}$, and Qing-Hu Chen$^{2,1,\dag}$}

\address{
$^{1}$ Department of Physics, Zhejiang University, Hangzhou 310027,
P. R. China\\
$^{2}$ Center for Statistical and Theoretical Condensed Matter
Physics, Zhejiang Normal University, Jinhua 321004, P. R. China }

\begin{abstract}
Intriguing  effects produced by  random percolative disorder in
two-dimensional Josephson-junction arrays are studied by means of
large-scale numerical  simulations. Using dynamic scaling analysis,
we evaluate  critical temperatures and critical exponents in high
accuracy. With the introduction of site-diluted disorder,  the
Kosterlitz-Thouless phase transition is eliminated and  evolves into
continuous phase transition with a power-law divergent correlation
length. Moreover, genuine depinning transition and the related creep
motion are studied, distinct types of creep motion for different
disordered systems are observed. Our results not only are in good
agreement with the recent experimental findings, but also shed some
light on the relevant phase transitions.
\end{abstract}


\maketitle

\section{Introduction}

Understanding the critical behavior of Josephson-junction arrays
(JJA's) with various disorders is always a challenging question and
has been intensely studied in recent years
\cite{harris}--\cite{yunepl}. However, the properties of different
phases and various phase transitions are not well understood.
Josephson-junction arrays
 gives an excellent realization to both  two-dimensional (2D)
XY model and granular High-$T_{c}$ superconductors \cite{lobb}. As
is well known that the pure JJA's undergoes celebrated
Kosterlitz-Thouless (KT) phase transition \cite{kt} from the
superconducting state to the normal state, this transition is driven
by the unbinding of thermally created topological defects. When the
disorder is introduced, the interplays between the periodic pinning
potential caused by the discreteness of the arrays, the repulsive
vortex-vortex interaction and the effects produced by the disorder
provide a rich physical picture.

In diluted JJA's, islands are randomly removed from the square
lattice. Since it is a representative model for realizing the
irregular JJA's systems, how the percolation influences the physical
properties of JJA's has attracted considerable attention
\cite{granato,harris,yun,Benakli2,Granato2}. Harris $et$ $al$
introduced random percolative disorder into Nb-Au-Nb
proximity-coupled junctions, the current-voltage ($I$-$V$)
characteristics were measured and the results demonstrated that the
only difference of the phase transition compared with that in ideal
JJA's system is the decrease of critical temperature, while the
phase transition still belong to the KT-type with the disorder
strength spanning from $p=0.7$ to $p=1.0$ (here $1-p$ is the
fraction of diluted sites) \cite{harris}. However, a recent
experimental study by Yun $et$ $al$ showed that the KT-type phase
transition in unfrustrated JJA's was eliminated due to the
introduction of site-diluted disorder \cite{yun}. Therefore, the
existence of the KT-type phase transition in site-diluted JJA's
remains a topic of controversy.

On the other hand, much attention has been paid to investigate the
zero-temperature depinning transition and the related
low-temperature creep motion  both
theoretically~\cite{Nattermann,Chauve,Blatter} and
numerically~\cite{Rosters,luo,olsson1} in a large variety of
physical problems, such as charge density waves~\cite{Nattermann},
random-field Ising model~\cite{Rosters}, and flux lines in type-II
superconductors~\cite{luo,olsson1}. Since the non-linear dynamic
response is a striking aspect,  there is increasing interest  in
these systems, especially in the flux lines of type-II
superconductors~\cite{luo,olsson1}. In a recent numerical study on
the three-dimensional glass states of flux lines, Arrhenius creep
motion and non-Arrhenius creep motion were observed with strong
collective pinning and weak collective pinning, respectively
~\cite{luo}.

In this work, we numerically investigate the finite-temperature
phase transition in site-diluted JJA's at different percolative
disorders, the zero-temperature depinning transition and the
low-temperature creep motion are also considered. The outline of
this paper is as follows. Section \ref{model} describes the model
and the numerical method briefly. In section \ref{sm}, we present
the simulation results, analyzing them by means of scaling analysis.
Section \ref{su} gives a short summary of the main conclusions.

\section{Model and simulation method}\label{model}

JJA's  can be described by the 2D XY model on a simple square
lattice, the Hamiltonian of which is \cite{olsson,chenhu}
\begin{equation}
H=-\sum\limits_{ <i,j>}J_{ij}\cos(\phi_{i}-\phi_{j}-A_{ij}),
\end{equation}
where the sum is over all nearest neighboring pairs on a 2D square
lattice, $J_{ij}$ denotes the strength of Josephson coupling between
site i and site j, $\phi_{i}$ specifies the phase of the
superconducting order parameter on site i, and
$A_{ij}=(2\pi/\Phi_{0})\int \mathbf{A} \cdot d \mathbf{l}$ is the
integral of magnetic vector potential from site i to site j  with
$\Phi_{0}$ the flux quantum. The direct sum of $A_{ij}$ around an
elementary plaquette is $2\pi f$, with $f$   the magnetic flux
penetrating each plaquette produced by the uniformly applied field,
measured in unit of $\Phi_{0}$. In this paper, $f=0$  and  $f=2/5$
are in focus. The system sizes are selected as $128\times128$ for
$f=0$ and  $100\times100$ for $f=2/5$, the finite size effects in
these sizes are negligible. Diluted sites are randomly selected,
then the nearest four bonds of which are removed from the lattice.
The same random-number seed is used to choose the diluted sites, the
percolative threshold concentration is about $0.592(1)$ for both
systems \cite{gebele}.

The resistivity-shunted-junction (RSJ) dynamics is incorporated in
the simulations, which can be described as \cite{chenhu,chentang}
\begin{equation}
{\frac{\sigma \hbar }{2e}}\sum_{j}(\dot{\phi _{i}}-\dot{\phi _{j}})=-{\frac{%
\partial H}{\partial \phi _{i}}}+J_{{\rm ex},i}-\sum_{j}\eta
_{ij},
\end{equation}
where $\sigma$ is the normal conductivity, $J_{ex,i}$ refers to the
external current,  $\eta_{ij}$ denotes the thermal noise current
with $<\eta_{ij}(t)>=0$ and  $<\eta_{ij}(t)\eta_{ij}(t')>=2\sigma
k_{B} T\delta(t-t^{'})$.

The fluctuating twist boundary condition  is applied in the $xy$
plane to maintain the current, thus the new phase angle
$\theta_{i}=\phi_{i}+r_{i}\cdot\Delta$ $(
\Delta=(\triangle_{x},\triangle_{y})$
 is the twist variable) is periodic in each
direction. In this way, supercurrent between site i and site j is
 given by $J^{s}_{i\rightarrow
j}=J_{ij}sin(\theta_{i}-\theta_{j}-A_{ij}-r_{ij}\cdot \Delta),$ and
the dynamics of $\Delta_{\alpha}$ can be written as

\begin{equation}
\dot{\Delta}_{\alpha}=\frac{1}{L^{2}}\sum\limits_{<i,j>\alpha}[J_{i\rightarrow
j}+\eta_{ij}]-I_{\alpha},
\end{equation}
where $\alpha$ denotes the $x$ or $y$ direction, the voltage drop in
$\alpha$ direction is $V=-L \dot{\Delta}_{\alpha}$. For convenience,
units are taken as $2e=\hbar=J_{0}=\sigma =k_{B}=1$ in the
following. Above equations can be solved efficiently by a
pseudo-spectral algorithm  due to the periodicity of phase in all
directions. The time stepping is done using a second-order
Runge-Kutta scheme with $\Delta t=0.05$. Our runs are typically
$(4-8)\times 10^{7}$ time steps and the latter half time steps are
for the measurements. The detailed procedure in the simulations was
described in Ref. \cite{chenhu,chentang}. In this work, a uniform
external current $I$ along $x$ direction is fed into the system.

Since RSJ simulations with direct numerical integrations of
stochastic equations of motion are very time-consuming, it is
practically difficult to perform any serious disorder averaging in
the present rather large systems. Our results are based on one
realization of disorder. For these very large samples, it is
expected to exist a good self-averaging effect, which is confirmed
by two additional simulations with different realizations of
disorder. This point is also supported by a  recent study of JJA's
by Um $et$ $al$ \cite{um}, they confirmed that a well-converged
disorder averaging for the measurement is not necessary, and
well-converged data for large systems at a single disorder
realization leads to a convincing result. In addition, simulations
with different initial states are performed and the results are
independent on the initial state we used. Actually, the hysteric
phenomenon is usually negligible in previous RSJ dynamical
simulations on JJA's \cite{um,Lim}. For these reasons, the results
from simulations with a unique initial state (random phases in this
work) are accurate and then convincing.

\section{RESULTS AND DISCUSSION}\label{sm}
\subsection{Finite temperature phase transition}

The $I$-$V$ characteristics are measured at various disorder
strengths and temperatures. At each temperature, we try to probe the
system at a current as low as possible. To check the method used in
this work, we investigate the $I$-$V$ characteristics for
$f=0,p=1.0$. As shown in Fig. 1(a), the slope of the $I$-$V$ curve
in log-log plot at the transition temperature $T_{c}\approx 0.894$
is equal to 3, demonstrating that the $I$-$V$ index jumps from 3 to
1, in consistent with the well-known fact that the pure JJA's
experiences a KT-type phase transition at $T_{c}\approx 0.894$.
Figs. 1(b) and (c)  show the $I$-$V$ traces at different percolative
disorders in unfrustrated JJA's, while Fig. 1(d) for $f=2/5,p=0.65$.
It is clear that, at lower temperatures, $R=V/I$ tends to zero as
the current decreases, which follows that there is a true
superconducting phase with zero linear resistivity.

It is crucial to use a powerful scaling method to analyze the
$I$-$V$ characteristics. In this paper, we  adopt the
Fisher-Fisher-Huse (FFH) dynamic scaling method, which provides an
excellent approach to  analyze the superconducting phase transition
\cite{FFH}. If the properly scaled $I$-$V$  curves collapse onto two
scaling curves above and bellow the transition temperature, a
continuous superconducting phase transition is ensured. Such a
method is widely used recently \cite{grt,yang}, the scaling form of
which in 2D is
\begin{equation}
V=I\xi^{-z}\psi_{\pm}(I\xi),
\end{equation}
where $\psi_{+(-)}(x)$ is the scaling function above (below)
$T_{c}$, $z$ is the dynamic exponent, $\xi$ is the correlation
length, and $V \sim I^{z+1}$ at $T=T_{c}$.

Assuming  that the transition is continuous and characterized by the
divergence of the characteristic length $\xi\sim|T-T_{c}|^{-\nu}$
and time scale $t\sim\xi^{z}$, FFH dynamic scaling takes the
following  form
\begin{equation}\label{eq:scv}
(V/I)|T-T_{c}|^{-z\nu}=\psi_{\pm}(I|T-T_{c}|^{-\nu}).
\end{equation}

On the other hand, a new scaling form is successfully adopted to
certify a KT-type phase transition in JJA's by \cite{holzer}
\begin{equation}\label{eq:sc2}
(I/T)(I/V)^{1/z}=P_{\pm}(I\xi/T),
\end{equation}
note that the Eq. (\ref{eq:sc2}) can be obtained directly from the
FFH dynamic scaling form after some simple algebra. The correlation
length of KT-type phase transition above $T_{c}$ is well defined as
$\xi\sim e^{({c/|T-T_{c}|})^{1/2}}$ and  Eq. (\ref{eq:sc2}) reads
\begin{equation}\label{eq:sckt}
(I/T)(I/V)^{1/z}=P_{+}(I e^{({c/|T-T_{c}|})^{1/2}}/T).
\end{equation}

\begin{figure}
\includegraphics[width=0.5\textwidth,height=10cm] {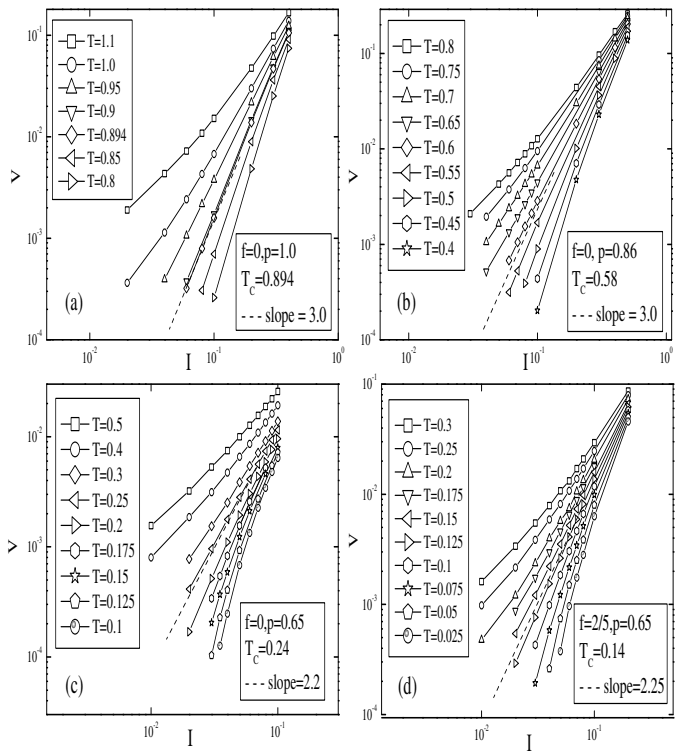}
\caption{\label{Figure Graph1} $I$-$V$  characteristics for
different frustrations and percolative disorders. The dash lines are
drawn to show where the phase transition occurs, the slopes of which
are equal to $z+1$, z is the dynamic exponent.  The transition
temperature and dynamic exponent  for (a) are well consistent with
the well-known result, $i.$ $e.$, $T_{c}=0.894$, $z=2.0$, for
(b),(c),(d) are well consistent with those determined by FFH dynamic
scaling analysis. Solid lines are just guide to eyes.}
\end{figure}

As shown in Fig. 2, using $T_{c}=0.24\pm0.01$, $z=1.2\pm0.02$ and
$\nu=1.0\pm0.02$, we get an excellent collapse for $f=0,p=0.65$
according to equation (\ref{eq:scv}). In addition,
 all the low-temperature $I$-$V$  curves   can be fitted to $V\sim
I$exp$(-(\alpha/I)^\mu)$ with $\mu=0.9\sim1.1$. These results
certify a continuous superconducting phase with long-rang phase
coherence. The critical temperature for such a strongly disordered
system is very close to that in 2D gauge glass model ($T=0.22$)
\cite{chen3}.

\begin{figure}
\includegraphics[width=0.4\textwidth,height=5.5cm]{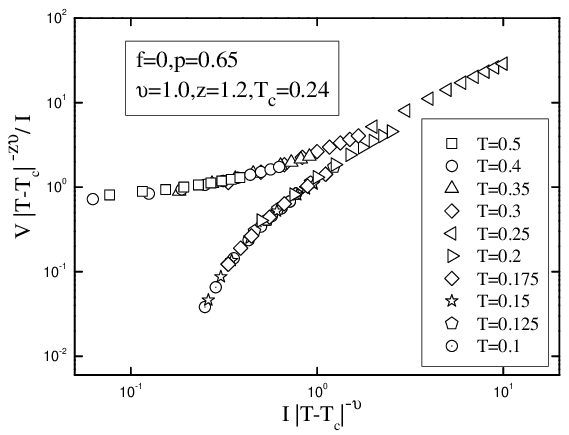}
\caption{\label{Figure Graph2}  Dynamic scaling of $I$-$V$  data at
various temperatures  according to equation (\ref{eq:scv}) for
$f=0,p=0.65$. }
\end{figure}

For $f=0,p=0.86$, firstly, we still adopt the scaling form in
equation (\ref{eq:scv}) to investigate the $I$-$V$ characteristics.
As displayed in Fig. 3, we get a good collapse for $T<T_{c}$ with
$T_{c}=0.58\pm0.01$, $z=2.0\pm0.01$ and $\nu=1.4\pm0.02$,
demonstrating a superconducting  phase with power-law divergent
correlation for $T<T_{c}$. Note that the collapse is bad for
$T>T_{c}$, indicating that the phase transition is not a completely
non-KT-type one. Next, we use the scaling form in equation
(\ref{eq:sckt}) to analyze the $I$-$V$ data above the critical
temperature. Interestingly, using $T_{c}=0.58$ and $z=2.0$
determined above, a good collapse for $T>T_{c}$ is achieved, which
is shown in Fig. 4. That is to say, the $I$-$V$ characteristics at
$T<T_{c}$ are like those of a continuous phase transition with
power-law divergent correlation length while at $T>T_{c}$ are like
those of KT-type phase transition, which are well consistent with
the recent experimental observations \cite{yun}. Therefore, by the
present model, we  recover the phenomena in experiments and give
some insight into the phase transition. More information on the
low-temperature phase calls for further equilibrium Monte Carlo
simulations as in Ref. \cite {Katzgraber}.

\begin{figure}
\includegraphics[width=0.4\textwidth,height=5.5cm]{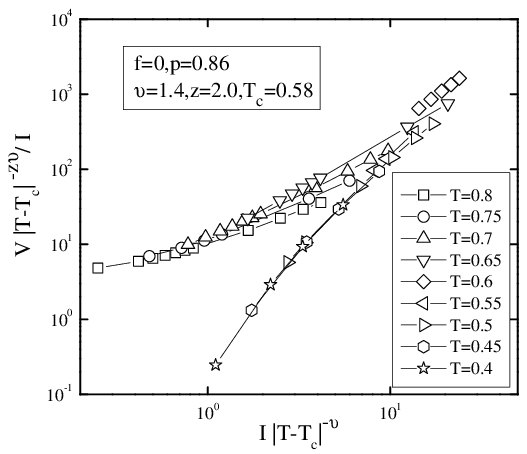}
\caption{\label{Figure Graph3}   Dynamic scaling of $I$-$V$ data at
various temperatures   according to equation (\ref{eq:scv}) for
$f=0,p=0.86, T<T_{c}$. Solid lines are just guide to eyes.}
\includegraphics[width=0.4\textwidth,height=5.5cm]{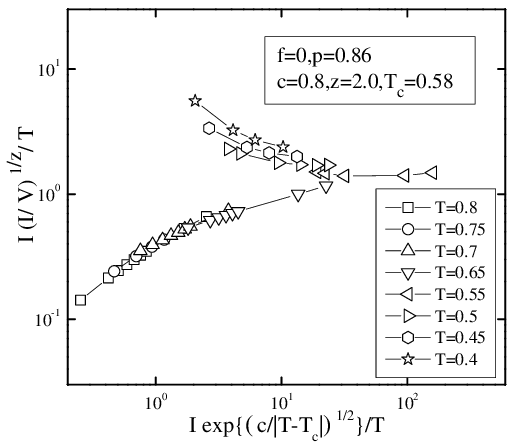}
\caption{\label{Figure Graph4}  Dynamic scaling of $I$-$V$  data at
various temperatures according to equation (\ref{eq:sckt}) for
$f=0,p=0.86, T>T_{c}$. Solid lines are just guide to eyes.}
\end{figure}

To make a comprehensive comparison with the experimental findings as
in Ref. \cite{yun}, we also investigated the finite-temperature
phase transition in frustrated JJA's ($f=2/5$) at a strong
site-diluted disorder ($p=0.65$). As shown in Fig. 5, a
superconducting phase transition with power-law divergent
correlation is clearly observed. As is well known, non-KT-type phase
transition  in frustrated systems is a natural result. However, it
is intriguing to see that in unfrustrated systems,  one may ask what
our results really imply and what is the mechanism for it.  It has
been revealed that in the presence of a strong random pinning which
is produced by random site dilutions, a breaking of ergodicity due
to large energy barrier against vortex motion may allow enough
vortices to experience a non-KT-type continuous transition
\cite{holme}.

\begin{figure}
\includegraphics[width=0.4\textwidth,height=5.5cm]{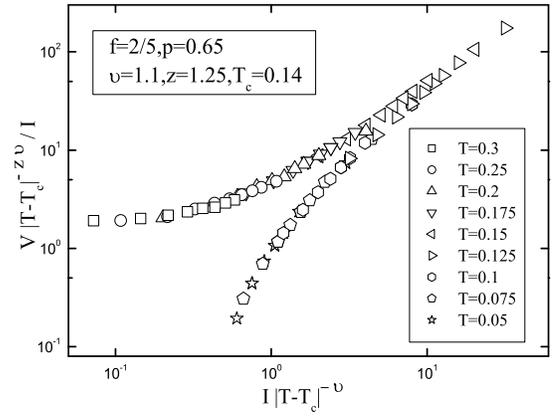}
\caption{\label{Figure Graph5}  Dynamic scaling of $I$-$V$  data at
various temperatures  according to equation (\ref{eq:scv}) for
$f=2/5,p=0.65$. }
\end{figure}

\begin{figure}
\includegraphics[width=0.35\textwidth,height=14cm]{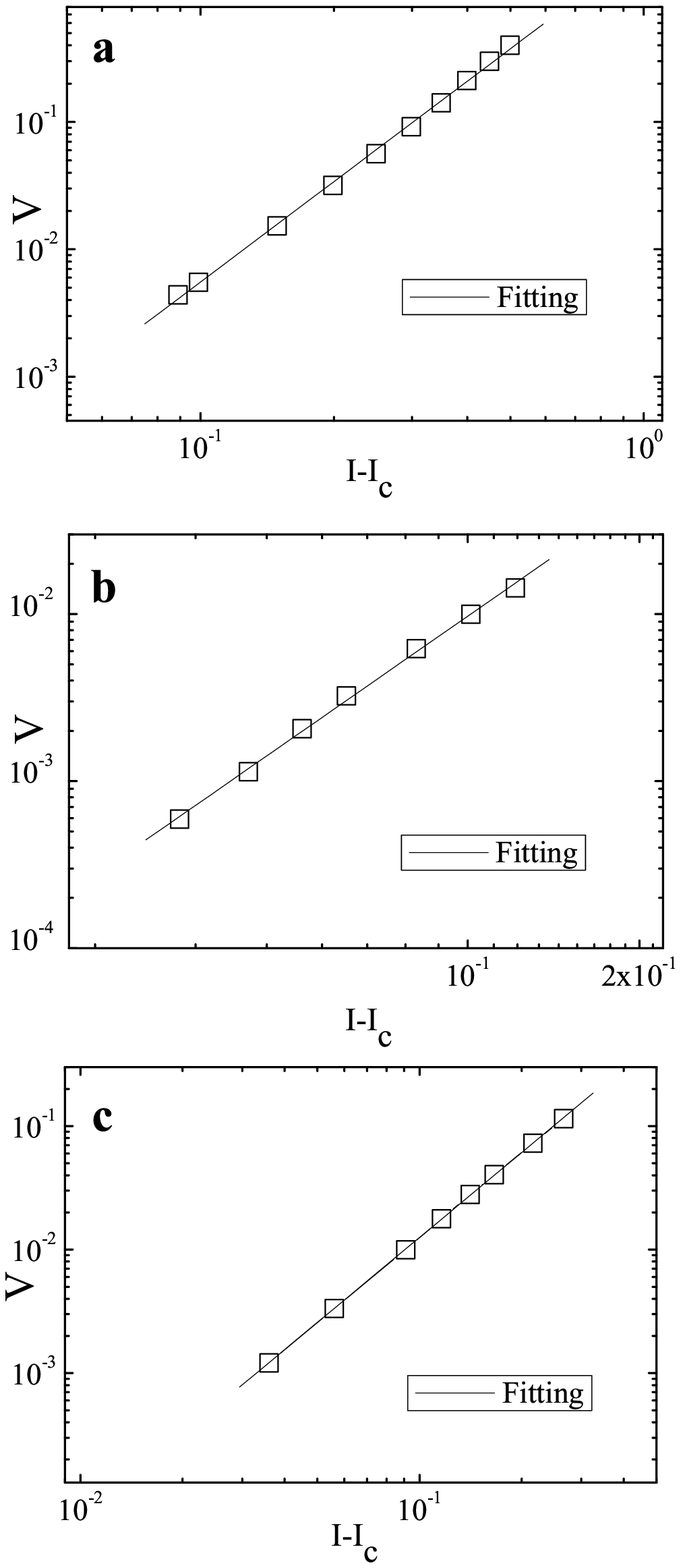}
\caption{\label{Figure Graph2} (a)$IV$ characteristics  for
$f=0,p=0.86$ with $I_{c}=0.302\pm0.005$, $\beta=2.62\pm0.1$. (b)$IV$
characteristics for $f=0,p=0.65$ with $I_{c}=0.039\pm0.001$,
$\beta=2.37\pm0.1$. (c)$IV$ characteristics for $f=2/5,p=0.65$ with
$I_{c}=0.035\pm0.002$, $\beta=2.27\pm0.05$}
\end{figure}

\begin{figure}
\includegraphics[width=0.35 \textwidth,height=12cm]{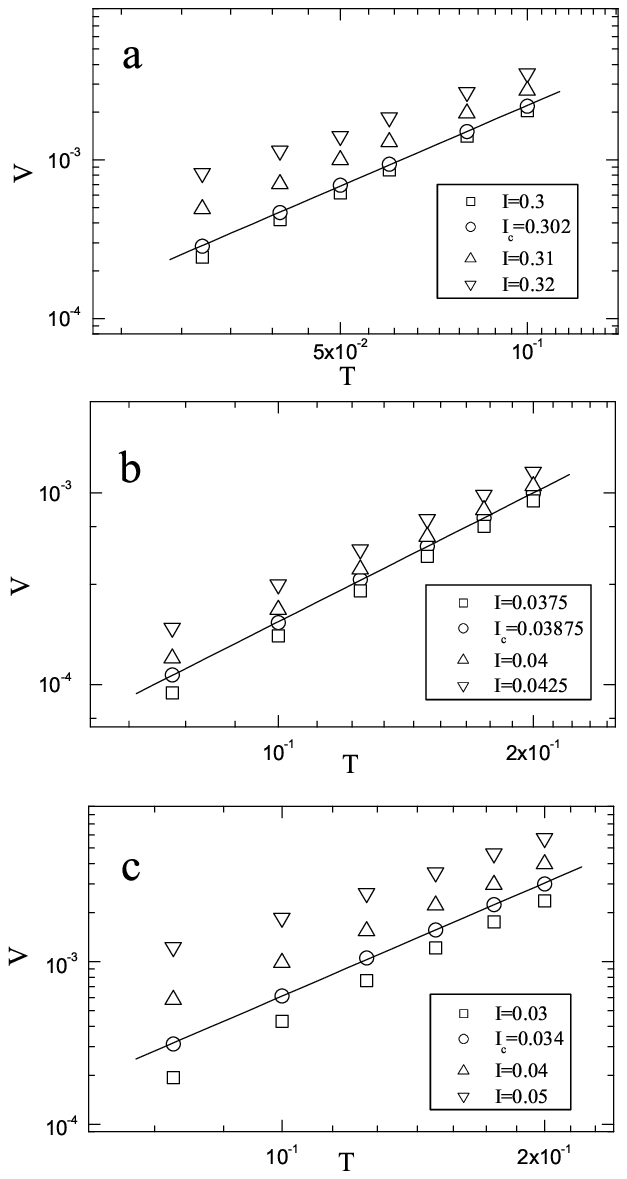}
\caption{\label{Figure Graph2} (a)LogV-LogT curves  for $f=0,p=0.86$
in the vicinity of $I_{c}$ with $I_{c}=0.302\pm0.001$,
$1/\delta=1.688\pm0.001$. (b)LogV-LogT curves for $f=0,p=0.65$ in
the vicinity of $I_{c}$ with $I_{c}=0.03875\pm0.0005$,
$1/\delta=2.24\pm0.02$. (c)LogV-LogT curves for $f=2/5,p=0.65$ in
the vicinity of $I_{c}$ with $I_{c}=0.034\pm0.001$,
$1/\delta=2.29\pm0.01$.}
\end{figure}

The systems considered in our work are site-diluted JJA's, which are
not the same  as bond-diluted JJA's in Ref.
\cite{Granato2,Benakli2}. The difference is, in bond-diluted systems
the diluted bonds are randomly removed, while in the site-diluted
systems, the diluted sites are randomly selected, then the nearest
four bonds around the selected sites are removed. Although the JJA's
in Ref. \cite{Granato2,Benakli2} and the present work are diluted in
different ways, it is interesting to note that  some of the obtained
exponents are very close, possibly due to the similar disorder
effect produced.

\subsection{Depinning transition and creep motion}
Next, we pay attention to the zero-temperature depinning transition
and the related low-temperature creep motion for the typical
site-diluted JJA's systems mentioned above. Depinning can be
described as a critical phenomenon with scaling law $V \sim
(I-I_{c})^{\beta}$, demonstrating a transition from a pinned state
below critical driving force $I_{c}$ to a sliding state above
$I_{c}$. The $(I-I_{c}).vs.V$ traces at $T=0$ for $f=0,p=0.86$;
$f=0,p=0.65$ and $f=2/5,p=0.65$ are displayed in Fig. 6,
linear-fittings of $Log(I-I_{c}).vs.LogV$ curves are also shown as
solid lines. As for $f=0,p=0.86$, the depinning exponent $\beta$ is
determined to be $2.62\pm0.1$ and the critical current $I_{c}$ is
$0.302\pm0.005$, while  for the cases $f=0,p=0.65$ and
$f=2/5,p=0.65$, the depinning exponents are evaluated to be
$2.37\pm0.1$ and $2.27\pm0.05$ with the critical currents
 $I_{c}=0.039\pm0.001$ and $I_{c}=0.035\pm0.002$,
respectively.

When the temperature increases slightly, creep motions can be
observed. In the low-temperature regime, the $I$-$V$ traces are
rounded near the zero-temperature critical current due to thermal
fluctuations. Fisher first suggested to map such a phenomenon for
the ferromagnet in magnetic field where the second-order phase
transition occurs ~\cite{fisher22}. This mapping was then extended
to the random-field Ising model ~\cite{Rosters} and the flux lines
in type-II superconductors ~\cite{luo}. For the flux  lines in
type-II superconductors, if the voltage is identified as the order
parameter, the current and the temperature are taken as the inverse
temperature and the field respectively, analogous to the
second-order phase transition in the ferromagnet, the voltage,
current and the temperature will satisfy the following scaling
ansatz ~\cite{luo,chen3}
\begin{equation}\label{eq:sca}
V(T,I)=T^{1/\delta}S[(1-I_{c}/I)T^{-1/\beta\delta}].
\end{equation}
The relation  $ V(T,I=I_{c})=S(0)T^{1/\delta}$ can be easily derived
at $I=I_c$, by which  the critical current $I_{c}$ and the critical
exponent $\delta$ can be determined through the linear fitting of
the $LogT-LogV$ curve at $I_{c}$. The $LogT-LogV$ curves are plotted
in Fig. 7(a) for $f=0,p=0.86$. We can observe that the critical
current is between 0.3 and 0.32. In order to locate the critical
current precisely, we calculate other values of voltage at current
within (0.3,0.32) with a current step 0.01 by quadratic
interpolation \cite{chen3}. Deviation of the $T$-$V$ curves from the
power law is calculated as the square deviations
$SD=\sum[V(T)-y(T)]^{2}$ between the temperature range we
calculated, here the functions $y(T)=C1T^{-C2}$ are obtained by
linear fitting of the $LogT-LogV$ curves. The current at which the
$SD$ is minimum is defined as the critical current. The critical
current is then determined to be $0.302\pm0.001$. Simultaneously, we
obtain the exponent $1/\delta =1.688\pm0.001$ from the slope of
$LogT-logV$ curve at $I_{c}=0.302$. The similar method is applied to
investigate the cases $f=0,p=0.65$ and $f=2/5,p=0.65$.  As shown in
Figs. 7 (b) and (c), the critical current $I_{c}$ and critical
exponent $1/\delta$ for $f=0,p=0.65$ are determined to be
$0.03875\pm0.0005$, $2.24\pm0.02$ respectively,  for $f=2/5,p=0.65$,
the result is $I_{c}=0.034\pm0.001$, $1/\delta=2.29\pm0.01$.
\begin{figure}
\includegraphics[width=0.4\textwidth,height=17cm]{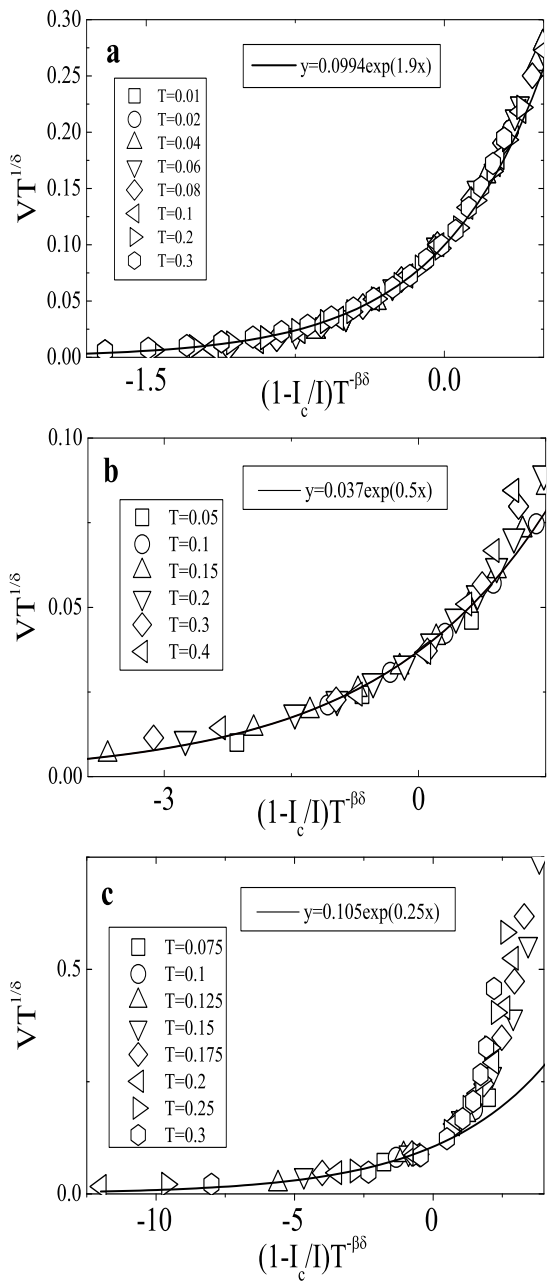}
\caption{\label{Figure Graph3} (a) Scaling plot for $f=0,p=0.86$
with $I_{c}=0.302$, $1/\delta=1.688$ and $\beta\delta=1.55$. (b)
Scaling plot for $f=0,p=0.65$ with $I_{c}=0.03875$, $1/\delta=2.24$
and $\beta\delta\approx1.0$. (c)Scaling plot for $f=2/5,p=0.65$ with
$I_{c}=0.034$, $1/\delta=2.29$ and $\beta\delta\approx1.0$.}
\end{figure}

We then draw the scaling plots according to Eq. 8. Using the one
parameter tuning of $\beta$, we get the best collapses of data to a
single scaling curve with $\beta=2.61\pm0.02$ and $2.28\pm0.02$ for
$f=0,p=0.86$ and $f=0,p=0.65$ in the regime $I \leq I_{c}$,
respectively, which are shown in Figs. 8(a) and (b). For
$f=0,p=0.86$, this curve can be fitted by $S(x)=0.0994exp(1.9x)$,
combined with the relation $\beta\delta=1.55$, suggesting a
non-Arrhenius creep motion. However, for the strongly site-diluted
system with $f=0,p=0.65$, the scaling curve  can be fitted by
$S(x)=0.037exp(0.5x)$, combined with the relation $\beta\delta
\approx 1.0$, indicative of an Arrhenius creep motion.
Interestingly,  as displayed in Fig. 8(c) for $f=2/5,p=0.65$, the
 exponent $ \beta $ is fitted to be $2.30\pm0.02$, which yields  $\beta \delta
\approx 1.0$. The scaling curve in the regime $I \leq I_{c}$ can be
fitted by $S(x)=0.105exp(0.25x)$.  These two combined facts suggest
an Arrhenius creep motion in this case.

It is worthwhile to note that both the finite-temperature phase
transition and the creep motion for strongly disordered JJA's
($p=0.65$) with and without frustration are very similar. The
$I$-$V$ curves in low temperature for all three cases can be
described by $V\propto T^{1/\delta} exp[
A(1-I_{c}/I)/T^{\beta\delta}]$, which is just  one of the main
characteristics of glass phases \cite{luo,chen3}. While the $I$-$V$
traces for  KT-type phases can be fitted to $V \propto I^{a}$.
Therefore, we have provided another evidence for the existence of
non-KT-type phases in the low-temperature regime for these three
cases ($f=0,p=0.86; f=0,p=0.65; f=2/5,p=0.65$).

\section{SUMMARY}\label{su}

To explore the properties of various phase transitions in
site-diluted JJA's, we have performed large scale simulations  at
two typical percolative strengths $p=0.86$ and $p=0.65$ as in a
recent experimental work \cite{yun}.  The RSJ dynamics was
incorporated in our work, from which we measured the $I$-$V$
characteristics at different temperatures. The critical temperature
of the finite-temperature phase transition was found to decrease as
the diluted sites increase. For $f=0,p=0.86$, the phase transition
is the combination of a KT-type transition and a continuous
transition with  power-law divergent correlation length. At strong
percolative disorder ($p=0.65$), the KT-type phase transition  in
pure JJA's is changed into a completely non-KT-type phase
transition, moreover, the finite-temperature phase transition for
frustrated JJA's is similar to that in  unfrustrated JJA's. All the
obtained dynamic exponents $z=a-1$, with $a$ the $I$-$V$ index at
the critical temperature, and all the static exponents fall in the
range of $\nu =(1.0, 2.0)$ usually observed at vortex-glass
transitions experimentally. Following table summarizes the critical
temperatures at different frustrations and  disorder strengths.

\begin{table}[h]
\caption{Summary of $T_{c}$.}
\begin{tabular}{p{3cm}  p{2cm} p{2cm} }
 \hline
 \hline
 & f=0 & f=2/5   \\
 \hline
 p=0.95 & 0.85(2) & 0.16(2)   \\
 p=0.86 & 0.58(1) & 0.13(1) \\
 p=0.7 & 0.27(2)  & 0.12(1)\\
 p=0.65& 0.24(1) & 0.14(1)\\
 \hline
 \hline
\end{tabular}

\end{table}

In a recent experiment, Yun et al \cite{yun} suggested a non-KT-type
phase transition in unfrustrated JJA's with site-diluted disorder
for the first time, however the nature of these phase transitions
and various phases is still in an intensive debate. Our results not
only recover the recent experimental findings \cite{yun}, but also
shed some light on the various phases. Non-KT-type
finite-temperature phase transition in site-diluted JJA's was
confirmed by the  scaling analysis. The different divergent
correlations at various disorder strengths were suggested, the
critical exponents were evaluated in high accuracy, which are
crucial for understanding such a critical phenomenon. Furthermore,
the results in this paper are not only useful for understanding the
site-diluted systems, but also useful for understanding the whole
class of disordered JJA's. For instance, the combination of two
different phase transitions may exist in other disordered JJA's
systems.

In addition, the zero-temperature depinning transition and the
low-temperature creep motion are also touched.  It is demonstrated
by the  scaling analysis  that the creep law for $f=0,p=0.86$ is
non-Arrhenius type while those for  $f=0,p=0.65$ and $f=2/5,p=0.65$
belong to the Arrhenius type. The evidence  of non-KT-type phase
transition can also be provided by this scaling analysis. It is
interesting to note that the non-Arrhenius type creep law for weak
disorder ($f=0,p=0.86$) is similar to that in three-dimensional flux
lines with a weak collective pinning \cite{luo}. The product of the
two exponents $1.55$  is also very close to $3/2$ determined in
Ref.\cite{luo}. For $f=0,p=0.65$ and $f=2/5,p=0.65$, the observed
Arrhenius type creep law is also similar to that in the glass states
of flux lines with a strong collective pinning as in Ref.
\cite{luo}. Future experimental work is needed to clarify this
observation.

\section{ACKNOWLEDGEMENTS}

This work was supported by National Natural Science Foundation of
China under Grant Nos. 10774128, PCSIRT (Grant No. IRT0754) in
University in China,  National Basic Research Program of China
(Grant Nos. 2006CB601003 and 2009CB929104), and Zhejiang Provincial
Natural Science Foundation under Grant No. Z7080203.

$^{\dag}$ Corresponding author. Email:qhchen@zju.edu.cn

\end{document}